\documentclass[11pt, a4paper, logo, twocolumn, copyright]{googledeepmind}

\pdftrailerid{redacted}

\makeatletter
\renewcommand\bibentry[1]{\nocite{#1}{\frenchspacing\@nameuse{BR@r@#1\@extra@b@citeb}}}
\makeatother

\usepackage{kantlipsum, lipsum}
\usepackage{dsfont}
\usepackage{gdm-colors}

\usepackage[authoryear, sort&compress, round]{natbib}

\graphicspath{{figures/}}

\title{Wanting to Be Understood Explains the Meta-Problem of Consciousness}

\correspondingauthor{chrisantha@google.com}

\paperurl{arxiv.org/abs/123}

\reportnumber{001} % Leave blank if n/a

\renewcommand{\today}{2025-06-11}

\author[1]{Chrisantha Fernando}
\author[1]{Dylan Banarse}
\author[1]{Simon Osindero}

\affil[1]{Google DeepMind}

\begin{abstract}
Because we are highly motivated to be understood, we created public external representations—mime, language, art—to externalise our inner states. We argue that such external representations are a pre-condition for access consciousness, the global availability of information for reasoning. Yet the bandwidth of access consciousness is tiny compared with the richness of `raw experience', so no external representation can reproduce that richness in full. Ordinarily an explanation of experience need only let an audience `grasp' the relevant pattern, not relive the phenomenon. But our drive to be understood, and our low level sensorimotor capacities for `grasping' so rich, that the demand for an explanation of the feel of experience cannot be “satisfactory”. That inflated epistemic demand (the preeminence of our expectation that we could be perfectly understood by another or ourselves) rather than an irreducible metaphysical gulf—keeps the hard problem of consciousness alive. But on the plus side, it seems we will simply never give up creating new ways to communicate and think about our experiences. In this view, to be consciously aware is to strive to have one's agency understood by oneself and others. \\ 
\end{abstract}

\begin{document}

\maketitle

This paper proposes a solution to the meta-problem of consciousness, which is summarised below by Andy Clark et al. 

\begin{quote}
The ‘hard problem of consciousness’ is the problem (Chalmers (1996)) of explaining how physical events give rise to the varieties of conscious phenomenal experience. The meta-problem of consciousness (Chalmers, 2018) is the problem of explaining why we think there is a hard problem in the first place. It is the problem of explaining why it is that some intelligent agents find themselves deeply puzzled by certain features of their own contact with the world - puzzled enough, in some cases, to announce the existence of a profound ‘explanatory gap’ between their best imaginable scientific grip upon how physical things work and the nature and origins of their own experience \citep{clark2019bayesing}.
\end{quote}

Solving the meta-problem is easier than solving the hard problem because it does not require explaining why phenomenal experience is like it is, but only to explain the things people say and do, i.e. why they consider the hard problem of consciousness a problem at all, e.g. why they are puzzled about qualia, about why red looks the way it looks. Chalmers categorises these intuitions as i. intuitions holding that consciousness is hard to explain, ii. intuitions that consciousness is non-physical, iii. intuitions that consciousness involves special first-person knowledge, iv. that it is hard to know the consciousness of other people, v. that someone else might be experiencing green when I am experiencing red, vi. and that consciousness makes life worth living \citep{chalmers2018meta}. Why would an inference machine such as the brain be lead to conclude that there are puzzling states with the hallmarks of `qualia', and have similar intuitions? \\

\cite{clark2019bayesing} propose an explanation for the meta-problem. They claim it is solved once we see the brain as a hierarchical Bayesian inference engine whose mid‐level posterior beliefs (e.g. “redness,” “pain”) are inferred by predictive processing of the raw sensory flux with exceptionally high precision and certainty, because this was of adaptive advantage, e.g. not letting us question that a predator's red beak was really there. \\

They then state that these mid‐level encodings, while highly certain, under‐determine their explanations, so advanced agents learn to hold them fixed and then vary their higher‐level explanations (e.g. “water fountain” vs. “vodka fountain” or “dream”). They write ``as the depth and reach of generative models increased we became aware that things are not always as they seem", and we were able to become puzzled by our explanations of mid-level certainties, e.g. mysterious intervening qualia because they ``enable us [to] better predict our own and other's future responses''. In doing so the agents infer a mysterious qualitative realm—qualia—and report an explanatory gap, even though these qualia are themselves just latent variables in the generative model, not raw data.

\section{Wanting To Be Understood}

We believe that Clark et al.’s account misses the social engine (the dependency on intersubjective and cultural dynamics) that first creates talk of “qualia.” Humans possess a species-unique urge to make inner states public and comparable—a motivation Tomasello terms shared intentionality \citep{tomasello2023differences}.
That impulse drives us to coin words and other external representations for our experiences; only once such shared representations exist can the very concept of qualia arise \citep{fernando2024origin}. \\

Consider the experience of profoundly deaf children, born to hearing parents, who are left in a language-less state. They are deeply troubled by their isolation. Oliver Sacks' describes a profoundly deaf eleven year old boy who entered school with no language \citep{sacks2009seeing}. ``I was partly reminded in a way of a nonnverbal animal, but no animal ever gave the feeling of yearning for language as Joseph did. Hughlings-Jackson, it came to me, once compared aphasics to dogs -- but dogs seem complete and contented in their languagelessness, whereas the aphasic has a tormenting sense of loss. And Joseph, too: he clearly had an anguished sense of something missing, a sense of his own crippledness and deficit.". (p 33). Before sign, Joseph could not grasp the idea of a question, no sense of the past, life lacked autobiographical and historical dimension, life existed only in the moment, in the present. But he had visual intelligence, could draw, could learn tic-tac-toe, but was completely literal, his thoughts were restricted to the immediate. Joseph later writes ``I saw cattle, horses, donkeys, pigs, dogs, cats, vegetables, houses, fields, grapevines, and after seeing all these things remembered them well.'' (p 37). When given the opportunity to learn sign language such children escape from their proposition-less present and concrete moment, and strive to join a mutually aware community of external representation users. When placed together they develop home sign, a primitive system lacking the full grammar of natural language but which permits mutual awareness \citep{goldin2020meeting}. This case powerfully illustrates that without external representations we lack the tools not only for mutual understanding but even for self-reflection on our own experiences.\\

Higher level explanations such as `qualia' are not entertained in the languageless state, even though individuals still perform hierarchical Bayesian inference on sensory data. However there is a strong drive for shared intensionality \citep{tomasello2023differences}. We suggest that humans evolved an intrinsic motivation for shared intensionality through the Baldwin effect \citep{Baldwin1896, jablonka2014evolution}. In the Late‐Pleistocene niche, small foraging bands faced a relentless diversity of cooperative communicative problems—signalling prey locations, dividing labour, tracking debts, and engaging in sophisticated forms of bush-craft requiring signalling of events displaced in space and time \citep{fernando2024origin}. Models of language evolution show that such culturally learned skills can become canalised via Baldwinian selection \citep{Lipowska2007,Podlipniak2017,Kolodny2018,Morgan2020}, while work on intrinsic motivation \citep{OudeyerKaplan2007} and on ``cognitive gadgets'' \citep{heyes2018cognitive} explains how a culturally installed drive to be mutually understood could itself become directly heritable. Yet, as \citet{Hrdy2009} and \citet{Burkart2009} emphasise, the most reliable source of daily selective pressure may not have been the coordination of adult hunters but the cooperative rearing of helpless infants in extended families.  The need for babies to elicit care from multiple alloparents—and for adults to tolerate and respond to non-kin neonates—would have powerfully favoured the evolution of gregarious, attention-soliciting, empathy-seeking minds, providing exactly the socio-emotional substrate on which Baldwin-style cultural selection for shared intentionality could build. In other words, we wanted to be understood by each other for the sake of it, to reach agreement about the meanings of external representations. We suggest it is this motivation for wanting to be understood, which results in us having the above intuitions about the hard problem of consciousness \citep{fernando2024origin}. \\

On the other hand, chimpanzees and other non-human animals show no sign of trying to make their private experiences publicly effable. Although they possess first-order metacognitive abilities—e.g. opting out of difficult memory tests or seeking additional information when uncertain \citep{hampton2001metacognition,call2001information}—they do not invent new symbols or vocalisations to convey the quality of those states to others. Great-ape calls remain tightly bound to immediate external events \citep{slocombe2005functionally}, and long-term field and laboratory studies find no cumulative cultural pressure to refine those calls into referential labels for inner states \citep{liebal2014primate,griebel2024emotional}. Even apes trained on artificial lexigrams use them mainly instrumentally and do not coin terms for novel feelings or perceptions \citep{savage1993language}. In short, non-humans monitor their own knowledge but show no motivation to share ineffable content. Humans, by contrast, play a “consciousness game’’ aimed at minimising this ineffability: we continually create new words, metaphors, and art forms to align others with our inner life—a species-unique drive for mutual awareness \citep{tomasello2008origins}. \\

The motivation to be understood plausibly underwrites the emergence of conversational heuristics such as Grice’s cooperative principle and its four maxims \citep{grice1975logic}. Psycholinguistic shows that speakers spontaneously tailor utterances to a listener’s knowledge state, revising descriptions until joint reference is achieved \citep{clark1986referring,brennan1996conceptual}. During successful dialogue, speaker and listener neural activity becomes temporally coupled in reward-sensitive regions (e.g.\ ventral striatum), when “on the same page’’ \citep{hasson2012brain,stephens2010speaker}. Computational models likewise benefit from explicit understanding rewards: multi-agent reinforcement learners that receive social-influence or mutual-information bonuses converge on more informative messages \citep{jaques2019social}. We therefore hypothesise a biologically grounded ``understanding-monitor'' whose feedback shapes conversational behaviour; Grice’s maxims may reflect the structure of this internal reward signal. \\

Whilst active inference is a drive to want to understand (personally) through the neural discovery of latent variables in a single brain \citep{friston2010free,friston2017active}, and it is perhaps sufficient to explain a private predictive insight, it is less clear whether it is sufficient to explain the need to externalize such insights. To explain that, we posit an intrinsic motivation to be understood by another—a species-unique urge for shared intentionality and cooperative communication \citep{tomasello2008origins,grice1975logic,heyes2018cognitive}. Evaluating whether one has indeed been understood then supplies a social feedback signal \citep{jaques2019social}\citep{fernando2025wanting}, driving individuals to cast private inferences into public form—mime, performance, art, and, ultimately, language—as theorised in accounts of the cultural evolution of external representations \citep{donald1991origin,gardenfors2004conceptual,galantucci2005emergence, fernando2024origin}. \\

\section{External Representations And Access Consciousness}

Clark et al. leave unanswered the question of how such high-level explanations as qualia—and indeed even the very term red—could have arisen in the first place. They have not shown that hierarchical Bayesian inference alone is sufficient to explain the emergence of language, although such claims have been sketched elsewhere \citep{friston2020generative}. Linguistic convention itself needs an evolutionary–cultural account. Consider Clark et al.’s own example: “The redness of the predator’s beak, likewise, should be processed fast and with high enough confidence to recruit immediate evasive action.” The attribution of the word redness already presupposes a non-trivial convention: colour terms vary cross-culturally and follow identifiable evolutionary trajectories \citep{berlin1969basic,kay2023world}. We cannot be certain that your redness is my redness; we share only an agreed label anchored in public use \citep{lupyan2016centrality}. Thus, even a seemingly “low-level” belief about colour is in fact a belief about a communicative convention that depends on an external representation (the word red) and on pragmatic norms governing how the term should be used in discourse. \\

A large body of work suggests that the emergence of access consciousness—the ability to hold, manipulate, and report selected contents for reasoning and control—was catalysed by the invention of external representations. Developmentally, children’s “private speech’’ and gesture function as scaffolded working memory; when these overt signs are experimentally suppressed, problem-solving and error-monitoring deteriorate, indicating that the external tokens themselves carry the load of conscious access \citep{vygotsky2012thought,fernyhough2004private}. In adults, writing, sketching, and token-moving reliably expand the capacity of short-term memory beyond the classic 4–7-item limit, a phenomenon Kirsh and Maglio dub “epistemic action’’ \citep{miller1956magical,kirsh1994epistemic}. Neurocognitive studies show that literacy and diagram use recruit fronto-parietal networks normally associated with the global workspace, effectively “outsourcing’’ portions of working memory to paper and screen \citep{baars2005global,dehaene2009reading}. Philosophically, the extended-mind and material-engagement theses argue that symbols in the public environment become non-neutral components of the cognitive system, underwriting the reportability that Ned Block identifies with access consciousness \citep{block1995confusion,clark1998extended,malafouris2013things}. Our own archaeological-computational analysis shows that each major expansion of representational media—from Acheulean marks to digital code—co-occides with new layers of metacognitive report and self-regulation, consistent with the view that external representations are the enabling technology of access consciousness \citep{fernando2024origin}. \\

\begin{figure*}[h!]
\includegraphics[width=17cm]{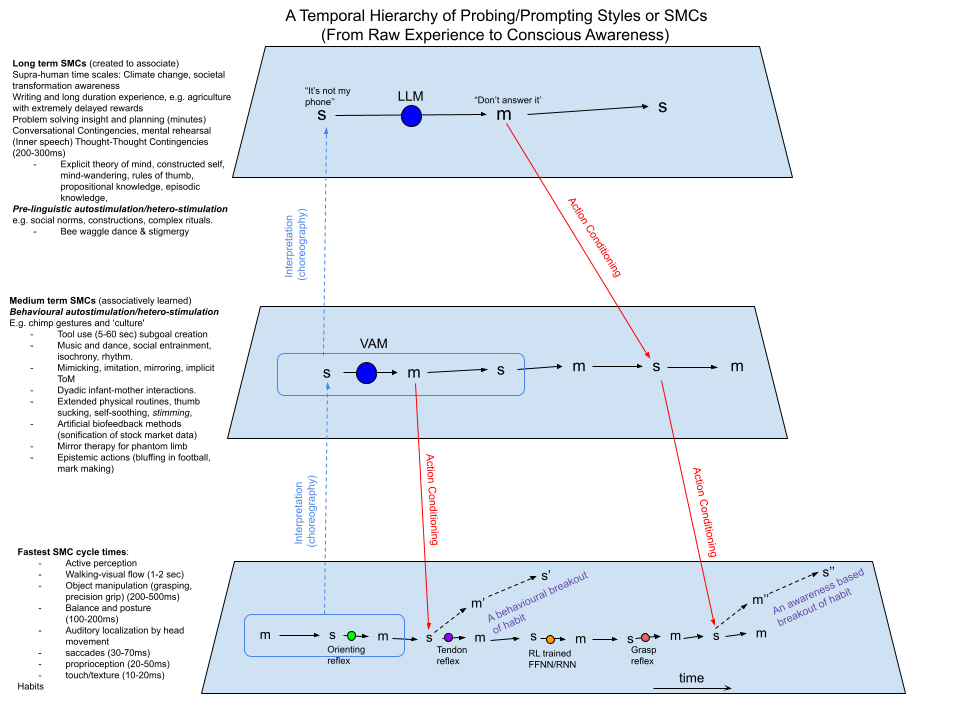}
\caption{Different time scales of sensorimotor contingency which includes autostimulatory probes/prompts. External representations are an advanced subset of what Dennett calls autostimulatory methods \citep{dennett1993consciousness}. These are themselves within a broader set of processes called sensorimotor contingencies (SMCs) which are shown here at different levels of description from fast acting (bottom) to slow acting (top). m = motor, s = sensory, and the arrows denote their contingencies. Bringing into awareness involves creating a new kind of SMC by autostimulation which allows control of previously habitual low level SMCs. For example two main ways to gain conscious access to autonomic functions are biofeedback training and providing emotionally evocative mental images (p67, \citep{baars2005global}). Biofeedback training always involves awareness. It is possible to gain control of alpha waves seen or heard in the EEG, or muscle activity when the sound of muscle fasciculations is played back to the agent (p101 \citep{baars2005global}).} 
\label{KindsOfProbe}
\end{figure*}

Viewed from an enactivist sensorimotor–contingency perspective, those very external symbols that scaffold access consciousness double as deliberate acts of autostimulation—self-generated probes we launch into our own perception–action loops to steer, stabilise, and implement thought. Figure~\ref{KindsOfProbe} illustrates different kinds of such broadly interpreted sensorimotor contingencies (SMCs) spanning multiple temporal scales. It aims to unify Dennett's concept of autostimulation \citep{dennett1993consciousness} within the framework of sensorimotor contingency, framing the broad concept of a probe as any kind of sensorimotor act. This includes ``thought'' as argued by \cite{vygotsky2012thought} is firstly a externalized behavioural act . Seen this way, there is a many-to-one mapping between long-term (thought-level) SMCs at the top and low-level (action-level) SMCs at the bottom \citep{dennett1993consciousness}. When a word is communicated, it cannot uniquely refer to the vast set of SMCs experienced at the lower level, a limitation echoed in predictive-processing models of language and perception \citep{clark2016surfing}. Probing processes associated with access consciousness thus occupy the higher tiers of the hierarchy, whereas those tied to phenomenal consciousness reside nearer the sensorimotor base \citep{baars2005global,noe2004action}\footnote{In a related approach, Vervaeke and colleagues argue that human cognition is organised around four irreducible “ways of knowing’’—\emph{propositional} (knowing–\textit{that}), \emph{procedural} (knowing–\textit{how}), \emph{perspectival} (knowing–\textit{what-it-is-like}\,/salience) and \emph{participatory} (knowing–\textit{with} and \textit{through} the world of shared affordances). Wisdom, they claim, arises when these modes are dynamically integrated by a meta-process they call \emph{relevance realisation}, the continual self-organisation that selects what matters at every scale of neural and behavioural activity \citep{VervaekeFerraro2013Wisdom}. In a more formal treatment, Vervaeke and Ferraro \citep{Vervaeke2012Relevance} locate each way of knowing in a distinctive memory system and normative “success criterion’’ (truth for propositional, skillful attunement for procedural, situational presence for perspectival, and connectedness/meaning for participatory) \citep{VervaekeFerraro2013Smart}.\\}. \\

Figure 1 also clarifies our departure with Dennett and other illusionists who argue that there is no gap between judging something to be pink and really seeming to experience pink—“the judgement is the whole show’’ (p.,364 \citealp{dennett1993consciousness}; cf.\ \citealp{cohen2011conscious,blackmore2002there}). Where we part company is over sufficiency: the propositional judgment alone does not exhaust the lived, qualitative texture of pink—for example, its felt warmth or salience against a green background. In short, the capacity to make the judgment is a necessary condition for the experience, but not a sufficient one. Block’s distinction between phenomenal and access consciousness captures this point: reportable, “accessed’’ content is only a thin slice of a much richer phenomenal field \citep{block1995confusion,block2007consciousness}. Sensorimotor-enactive theories likewise emphasise that experiential quality is constituted by the pattern of sensorimotor contingencies engaged during embodied probing \citep{o2001sensorimotor,noe2004action}. Probing-by-doing (eye saccades, head movements, tactile exploration) yields a high-bandwidth, parallel stream; probing-by-thinking is a slow, serial, low-bandwidth reconstruction. The sense of ineffable qualia arises precisely because the latter cannot recapitulate the information-rich former—a point echoed in empirical work showing massive perceptual detail outside the focus of report \citep{landman2003large,lamme2004separate}. In short, the “feel’’ of an English breakfast contains more raw information than the entire lexicon could encode; linguistic judgement can point to that feel but never fully substitute for it. \\

% The motivation to be understood explains the origin of socially constructed heuristics such as Grice’s cooperative ``Principle'' and its four accompanying maxims \citep{grice1975logic}. His idea is that in conversation we tacitly agree to make our contributions as helpful—and thus as understandable—as possible. The cooperative principle assumes both speaker and listener aim to communicate effectively and work together to do so. This entails a maxim of Quality: be truthful, the maxim of quantity: be as informative as is required, the maxim of Relation (Relevance): be relevant, and the maxim of manner: be clear (avoid obscurity of expression). Grice however, does not derive this principle or explain its origin. We present the empirical hypothesis that a mechanism for evaluating whether one has been understood exists, and that Grice's principles may derive from the form of this intrinsic reward. \\

\section{Bandwidth Differences} 

David Marr estimated that the human optic nerve conveys on the order of 10 Mb per second of raw visual data—roughly one million ganglion‐cell axons signalling at ~10 spikes per second during natural viewing \citep{marr1982vision}. By contrast, conscious report seems to be limited to only a few hundred bits per second, as inferred from discrimination thresholds, reaction times, and the psychological-refractory-period paradigm \citep{koch2007attention,vanrullen2003percept}. Visual working memory is tighter still—typically $\leq$ 4–7 individuated items or their resource equivalent \citep{miller1956magical,bays2008dynamic,vogel2004neural}—and when gaze shifts are prevented, accessible detail falls to just a handful of bits \citep{o2001sensorimotor}. Nevertheless, Landman et al.\ showed that observers retain far richer scene information than they can voluntarily report, suggesting a dissociation between phenomenal capacity and access capacity \citep{landman2003large}. Block leverages such findings to distinguish phenomenal consciousness, which may “overflow’’ any bottleneck, from access consciousness, which is constrained by working-memory and decision circuitry \citep{block2007consciousness}. \\

We will never be able to predict an experience of red perfectly even with practical, sensorimotor ``action probing'', let alone convey it to others through purely propositional ``thought probing’’ \citep{noe2004action,clark2015surfing}. ``The world of experience must be greatly simplified and generalized before it can be translated into symbols’’ (p.,6 \citealp{vygotsky2012thought}). This very inability to define and communicate phenomenal experience exhaustively—highlighted in Jackson’s Mary argument \citep{jackson1986mary} and in Nagel’s “what-it-is-like’’ challenge \citep{nagel1974like}—lies at the heart of our fascination with qualia. Contemporary work in neuroscience underscores the point: cortical colour coding is highly individualised and context dependent, such that ostensibly identical stimuli elicit idiosyncratic neural patterns across observers \citep{brouwer2009representations,bannert2013decoding}. The lingering unease we feel is thus the unease of an agent perpetually trying—and necessarily failing—to compress a high-dimensional sensorimotor manifold into public symbols in order to minimise mutual prediction error \citep{friston2010free}. We may be mystified about qualia not because experience is non-physical, but because any external representation (word, swatch, code) under-samples what it is meant to denote, such that our low-level experiences cannot be re-presented faithfully other than by replicating the full set of SMCs involved in having them. \\

A parallel line of evidence for the cognitive “bandwidth ceiling’’ comes from reading speeds which shows adults read silently at a mean rate of only 238 words min$^{-1}$ for non-fiction and 260 words min$^{-1}$ for fiction \citep{Brysbaert2019ReadingRate}. At roughly 4 words s$^{-1}$, and assuming six eight-bit characters per word, this corresponds to 190 raw bits s$^{-1}$—two orders of magnitude below the 10 Mb s$^{-1}$ optic-nerve figure. Eye-movement work shows why: each content word attracts a fixation of 200–250 ms and the perceptual span rarely exceeds 15 letters, making sustained rates above 300 wpm impossible without loss of comprehension \citep{Rayner1998EyeMovements}. Computational modelling confirms that lexical access and syntactic integration must proceed in a largely serial fashion; in Just and Carpenter’s classic model, conceptual propositions are integrated at a pace of only a few dozen bits per second despite much faster visual inflow \citep{JustCarpenter1980ReadingTheory}. Thus, the linguistic stream of thought is intrinsically narrow, dovetailing with psychophysical and dual-task estimates of conscious access capacity.\\

\section{Puzzlement}

So far we have argued that humans are intrinsically motivated to be understood—to share phenomenal experience—and therefore invent external representations that furnish access consciousness, the public, reportable layer of thought and control \citep{Block1995,fernando2024origin}. We now suggest that our puzzlement over the hard problem does not stem from the kind of low-variance perceptual certainties Clark et al. emphasise, but from the ineffable complexity of lived experience that eludes any access code. Words like red compress the vast, high-bandwidth sensorimotor contingencies (tens of megabits per second flowing through the optic nerve \citep{marr1982vision}) into a handful of phonemes—well below the few-hundred-bits-per-second limit of conscious report \citep{miller1956magical,koch2007attention}. O’Regan and Noë’s sensorimotor-contingency (SMC) theory makes this explicit: perceptual quality inheres in the skilful mastery of law-like relations between movement and sensory change \citep{o2001sensorimotor,noe2004action}. Because any linguistic token maps one‐to‐many onto the sprawling space of such contingencies, we are left with a felt “explanatory gap”—a modern echo of Nagel’s “what it is like” problem \citep{nagel1974like}, of Chalmers’ claim that experience “outruns” physical description \citep{chalmers1996conscious}, and of Wittgenstein’s private-language argument that sensation terms get their meaning only in public use \citep{wittgenstein1953}. The hard problem, on this view, arises from the social discomfort produced by chronic information loss whenever we translate the richness of experience into the thin channel of public symbols \citep{landman2003large}. \\

% Any explanation involves two components: a socially agreed upon description of what is to be explained (the explicandum), and a similarly agreed upon framework for what constitutes a valid explanation. We claim that the hard problem exists (and will always exist) because we cannot ever agree on the phenomenon that is to be explained, e.g. our experience of red. We have already seen we are motivated socially to reach such agreements that de-privatize previously private experience. We call this the struggle for mutual awareness. The very endeavour to produce social explanations, which involves a. agreeing an a referent to explain, and b. explaining it, is a socially constructed act of considerable complexity, and it is the motivation to engage in such social acts which must be explained in order to understand why we believe that a hard problem of consciousness exists. \\ 

Any explanation involves two components: a socially agreed-upon description of what is to be explained (the explicandum) and a similarly agreed-upon framework for what counts as a valid explanation \citep{lombrozo2012science,salmon1984scientific}. We claim that the hard problem exists (and will always exist) because we can never fully agree on the phenomenon to be explained—e.g.\ “our experience of red’’—since those experiences are privately instantiated yet only publicly discussable through conventional symbols \citep{berger1966social,clark1986referring}. We have already seen that humans are motivated to forge such agreements in order to de-privatize experience; we call this the struggle for mutual awareness \citep{tomasello2008origins}. The very enterprise of producing social explanations—(a) agreeing on a referent and (b) articulating an acceptable account—is itself a socially constructed, norm-governed practice \citep{brandom1994making,kuhn1962structure}. Explaining why we feel a “hard problem’’ therefore requires explaining the motivation to engage in these practices in the first place. \

% In other words, we contend that the hard problem exists due to our intolerance (unique among species) of the inevitable loss of information that a socially agreed upon explanation of sensorimotor experience entails \citep{mcgregor2005levels} \citep{o2001sensorimotor}. Agents that are driven to reach shared and accurate references to each other's experiences will inevitably be unsatisfied with the fact that a low dimensional external representation system such as language or even painting or music can never capture the informational richness of their high dimensional irreducibly complex private raw experiences. \\

In other words, we contend that the hard problem exists due to our frustration (unique among species) of the inevitable loss of information that a socially agreed-upon explanation of sensorimotor experience entails \citep{mcgregor2005levels,o2001sensorimotor}. Agents that are driven to reach shared and accurate references to each other’s experiences will inevitably be unsatisfied with the fact that a low-dimensional external-representation system—language, painting, music—can never capture the informational richness of their high-bandwidth, irreducibly complex private experiences. \

% "We have no independent language for describing phenomenal qualities. As we have seen, there is something ineffable about them" (p 22 \citep{chalmers1997conscious}, see also \citet{jakab2000ineffability}). "The problem of consciousness (phenonomal experience) cannot be spirited away on purely verbal grounds" (p24, \citep{chalmers1997conscious}). This is precisely the point. We propose that the feeling of ineffability about the phenomenal arises because of the inevitable leftover, the residue of our sensorimotor stream, that cannot be accurately predicted by us better with words, but which we attend to in order to help us build better shared external representations, e.g. words for the nuanced varieties of "red". Language is a culturally evolved set of external representations we have invented to help reduce our uncertainty of each other's experiences. But the disagreeable fact is, however many words we invent to help us know each other's worlds better, we will never completely capture all the properties of our sensorimotor contingencies (raw experiences) of red. \\

"We have no independent language for describing phenomenal qualities. As we have seen, there is something ineffable about them" (p. 22 \citep{chalmers1997conscious}; see also \citealp{jakab2000ineffability}). "The problem of consciousness (phenomenal experience) cannot be spirited away on purely verbal grounds" (p. 24 \citep{chalmers1997conscious}). This is precisely the point. We propose that the feeling of ineffability about the phenomenal arises because of the inevitable leftover—the residue of our sensorimotor stream—that cannot be accurately predicted or compressed by words, but which we nonetheless attend to in order to build better shared external representations (e.g.\ new terms for nuanced varieties of “red’’) \citep{o2001sensorimotor,landman2003large}. Language itself is a culturally evolved set of external representations for reducing uncertainty about one another’s experience \citep{lupyan2016centrality}. Yet however many words we coin, we can never fully capture the fine‐grained structure of our sensorimotor contingencies—colour being a classic case, where lexical categories vary across cultures despite common perceptual substrates \citep{berlin1969basic,kay2011world}. Hence the persistent sense that private experience “overflows’’ public description, a point already implicit in Jackson’s “Mary’’ argument and Wittgenstein’s reflections on the limits of language \citep{jackson1986mary,wittgenstein1953}. \\

\section{The Cultural Quest for Perfect Mutual Understanding} 

Talk of the hard problem of consciousness is best read as one culturally specific—yet globally recurring—expression of a deeper human project: the urge to externalise experience so that it can be shared, coordinated, and contested. Across traditions, thinkers have grappled with the puzzle of how private consciousness can be rendered mutually intelligible. Mahāyāna Buddhism, for example, frames awareness as an enacted, interdependent process rather than an isolated “inner light’’—a view Thompson links to contemporary enactivism \citep{thompson2014waking}. Classical contemplative texts describe enlightenment as the moment when the split between subject and object collapses and one “knows and sees things as they really are” (yathā-bhūta-ñāṇadassana) \citep[p.~339]{thompson2014waking}. The Buddha makes the point tersely: “Whoever sees dependent origination sees the Dhamma; whoever sees the Dhamma sees dependent origination” \citep[SN 12.20]{Bodhi2000}. Zen master Dōgen echoes the same ideal of complete co-presence: “To study the self is to forget the self; to forget the self is to be enlightened by all things” \citep[p.~70]{Dogen2004}. Far from privileging a lone observer, Buddhism casts awakening as the collapse of self–other boundaries into a field of mutual, compassionate awareness. Classical Nyāya philosophers in India analysed śabda (testimonial knowledge) as an indispensable bridge between first-person experience and communal truth \citep{chakrabarti1994indian}. In sub-Saharan Africa, the ethic of ubuntu—“a person is a person through other persons’’—conceptualises selfhood as something that comes into being only when recognised by a community \citep{mbiti1969african}. Even modern Arabic Bedouin poetry, as analysed by Abu-Lughod, treats emotion terms as public performances of honour aimed at aligning audience and speaker \citep{abu1986veiled}.\\

From the \emph{Tower of Babel}, where a single language once united humankind before being “confounded” \citep{Genesis11}, to Aristophanes’ myth in Plato’s \emph{Symposium}, in which lovers literally seek to fuse their severed halves \citep[189c–193d]{PlatoSymposium}, stories across cultures picture an impossible return to perfect mutual understanding. In Apuleius’ tale of \emph{Psyche and Eros} the heroine endures trials to gaze openly on her divine husband and attain face-to-face knowledge \citep[Met.~IV–VI]{ApuleiusMetamorphoses}; the twin heroes Castor and Pollux bargain to share life and death so that neither brother is ever alone \citep{PindarNemean10}. China’s Qixi legend of \emph{Niilang and Zhinu} sets the lovers on opposite sides of the Milky Way, permitted to meet only when birds bridge the gap each year \citep{ShanHaiJing}. Vedic hymn X.125 personifies Speech (Vac/Savitar) as the power that “binds the people with one thought and one word” \citep{RigVeda10_125}. A southern-African Ubuntu creation story recalls a time when people felt each other’s hearts directly until selfishness shattered their empathic unity \citep[chap.~2]{Mbiti1969}. In Norse lore, Odin steals the mead of poetry from the giants so that any drinker can voice thoughts flawlessly to any listener \citep{SnorriSkaldskaparmal}.\\

The global fascination with such myths reflects a species-wide urge to overcome the chronic information loss that occurs whenever rich, private experience is squeezed into low-bandwidth public symbols. Because humans are deeply motivated to have their inner states truly grasped by others, we imagine paradises of unmediated communion and lament their loss, spinning tales that dramatise both the dream of total transparency and the tragic conditions—multiple languages, mortal bodies, cosmic separations—that keep it just out of reach.
Western philosophy inherits the same impulse. Geertz’s interpretive anthropology casts culture as a web of public symbols spun to make subjective states jointly legible \citep{geertz1973interpretation}. Markus and Kitayama’s cross-cultural psychology shows that many societies emphasise relational rather than independent selves, valuing harmony and mutual understanding over solitary introspection \citep{markus1991culture}. Rosenthal’s critique of Chalmers points to the fact that intuitions about ineffable qualia are less likely to be robust once one samples non-WEIRD populations \citep{rosenthal2019chalmers}. And Jaynes’s historiography suggests that only with the rise of complex social coordination did people begin explicitly thematising an inner “mind’’ separable from public utterance \citep{jaynes2000origin}.\\

Attachment research shows that humans are wired from the outset to seek attuned, two-way understanding. Bowlby’s ethological theory treats the infant’s protest at separation as an evolved signal that only subsides when the caregiver “gets” what the child needs \citep{Bowlby1969}. In Ainsworth’s \emph{Strange Situation}, securely attached babies use the caregiver as a “secure base” and display joy when the adult accurately reads their cues, whereas insecure dyads are marked by misunderstood bids for contact \citep{Ainsworth1978}. The still-face paradigm shows that when a parent freezes into an unresponsive mask, infants become visibly distressed and work feverishly to re-establish mutual engagement \citep{Tronick1978StillFace}. 

By the end of the first year, infants initiate joint-attention episodes—pointing, showing, and alternating gaze—not for instrumental gain but to share experience itself \citep{Tomasello1995}. Fonagy’s mentalization model argues that such “being seen as seeing” scaffolds a lifelong capacity to recognise minds in oneself and others; breakdowns in this process predict borderline personality disorder and other relational pathologies \citep{Fonagy2002}. Clinical evidence converges: autism is now framed partly as a reduction in the intrinsic \emph{social-motivation} to be understood and to understand \citep{Chevallier2012}. Psychoanalytic accounts likewise cast selfhood as a co-creation: Stern’s “affect attunement” and Benjamin’s “mutual recognition” both hinge on the infant’s expectation that another will mirror and validate felt states \citep{Stern1985,Benjamin1995}. Together these lines of work converge on a single developmental fact: thriving minds are minds that have been met—accurately, contingently, and mutually—by other minds.\\

Chalmers’ ``philosophical zombie” illustrates that we can imagine a creature whose outward behaviour is indistinguishable from ours yet whose inner life is utterly void of qualia \citep{chalmers1996conscious}\footnote{The word \emph{zombi/zonbi} travelled from West-Central African cosmologies (Kikongo \emph{nzambi}) to the French colony of Saint-Domingue with enslaved Africans, but its meaning was recast in plantation life. Death was thought to release the soul so it could “walk back” to Africa (“Guinea”); a bokor who captured the soul and reanimated the corpse created a \emph{zonbi}—a body condemned to labour with no will of its own. Ethnologist Alfred Métraux already heard Haitians describe a zombie’s existence as “the perfect continuation of slavery beyond the grave” \citep{Metraux1959Voodoo}. Lauro’s  study calls the figure “the limit-case of the human subject turned commodity,” forged in the crucible of New-World slavery \citep{Lauro2015PreHistory}. Hoermann shows how Haitian-Revolution pamphleteers and later writers equated colonial domination with a “general zombification,” a fate worse than death because it annihilated political agency while keeping the body alive \citep{Hoermann2016Figures}. Field evidence collected by Davis confirms that rural informants still fear not dying but having the \emph{ti bon ange} (little good soul) stolen, trapping them in servitude “for all eternity” \citep{Davis1983Zombie}. Thus, the zombie myth encodes a relational trauma—a refusal by the master to recognise the slave’s personhood—projected into the afterlife: an ultimate, intersubjective loss of agency.   \\

From the mid-twentieth century onward the zombie myth gradually lost its explicitly colonial framing and was retooled in Euro-American popular culture as a parable of \emph{private} psychic collapse. George Romero’s films recast the Haitian slave–corpse as an anonymous, affect-less consumer, literalising late-capitalist alienation—a shift cultural critics dub the “zombie renaissance’’ \citep{Bishop2009DeadMan}. Lauro and Embry characterise the post-1968 zombie as “a body without a self,’’ a carrier of advanced-capitalist logics in which persons become interchangeable nodes on global value chains \citep{LauroEmbry2008ZombieManifesto}. Surveying the trope’s further evolution, Christie and Lauro show that twenty-first-century texts push this logic inward: the modern zombie evacuates the \emph{individual} of experiential depth, turning consciousness itself into a vacant sensorimotor loop \citep{ChristieLauro2011BetterOffDead}. Literary fiction echoes the move; Colson Whitehead’s \emph{Zone One} centres on survivors whose defining feature is a hollowed, “neutral-grey’’ interiority, the undead as emblem of derealised subjectivity \citep{Swanson2014ZoneOne}. The migration from collective loss of agency to solitary loss of \emph{felt} experience thus mirrors a broader Western drift from relational selfhood to privatised interiority, revealing how the zombie figure continually adapts to whichever form of de-humanisation most worries the culture that consumes it. \\}. \\

The ease with which this thought-experiment comes to mind is, we suggest, rooted in our earliest experience of affective contingency. In the still-face paradigm, infants first engage in normal face-to-face play, then confront a suddenly emotionless caregiver; within seconds they become distressed and attempt to re-engage, then withdraw in despair when reciprocity fails \citep{Tronick1978StillFace}. The experiment dramatises a primal fear: the other’s body may be present, yet the expected feeling of “being met” is absent. \\

A comparable but more extreme disjunction appears in Capgras syndrome. Neuro-cognitive models locate the delusion in damage to pathways linking the fusiform face area with limbic evaluative systems: visual recognition of a loved one is intact, but the autonomic “familiarity” signal is missing, so the patient concludes the person is an impostor \citep{EllisYoung1990DelusionalMisidentifications,Hirstein1997Capgras}. The body behaves normally (the patient greets mother) while subjective affect is null—an experiential “zombie-other.” Depersonalisation and related dissociative disorders reveal the same mechanism turned inward. Functional imaging and psychophysiology show diminished limbic activation and blunted autonomic responses to emotional stimuli; patients describe themselves as robotic, numb, or “already dead” \citep{SierraBerrios1998Depersonalization}. Epidemiological work links these states to childhood interpersonal trauma, in which expected contingent caregiving was repeatedly violated \citep{Simeon2001TraumaDPD}.\\

Developmental neuroscience offers a bridge from infancy to adult psychopathology. Chronic mis-attunement in the first years disrupts right-hemisphere fronto-limbic circuits that support emotional resonance \citep{Schore2001RelationalTrauma}. Attachment researchers trace a pathway from disorganised early attachment to pathological dissociation, arguing that the mind defends itself by severing the link between representation and affect \citep{Liotti2004TraumaDissociation}. Thus, the capacity to conceive of zombies is not mere philosophical fancy: it echoes a well-documented biopsychosocial pattern in which behaviour is uncoupled from feeling, whether projected onto others (Capgras) or experienced in oneself (dissociation).\\

Taken together, these strands support our claim that the hard-problem discourse persists because it is a historically situated outgrowth of a species-wide drive to be understood—a drive that can manifest as harmony, testimony, ubuntu, or modern talk of “qualia,” but whose root is the same: minimising the gulf between private experience and shared representation. Tracing our fascination with “qualia” to a deep, species-wide need for mutual understanding explains why the hard-problem intuition feels so urgent, but it does not touch the logic of the hard problem itself. A genealogical story about our motives is a psychological–cultural account; Chalmers’ gap is a metaphysical claim that even a complete physical description leaves phenomenal facts under-determined. Showing that people are driven to articulate an explanatory gap therefore neither proves nor disproves that such a gap exists—it simply clarifies why we keep asking the question, which is why the focus here is on an explanation of the meta-problem.\\

\section{The Algorithmic Basis of Wanting to Be Understood} 

If one accepts that we have such an intrinsic motivation for wanting to be understood, a research program which naturally follows is how such in intrinsic reward could be implemented in a multiagent reinforcement learning setting, resulting in the invention of new external representations which could mediate open-ended access consciousness? We have shown recently in experiments modelling primary intersubjectivity in the perceptual crossing paradigm that avatars wish to spend time with each other (in preference to with inanimate objects) when they are motivated to both influence and be impressionable (influenced by) the other. Their reward is to the mutual information between their acts and the others subsequent acts, and vice versa \citep{fernando2025wanting}. This mutual social reward allows avatars to cooperate in games that would otherwise not have been cooperatively stable. \\

Of course a more significant challenge is to understand intrinsic motivations for secondary intersubjectivity. Recent progress in \emph{mechanistic} and \emph{concept-based} interpretability shows how a learning system can be \emph{directly rewarded} for exposing its internal reasoning about other things in the world in a form that another agent (human or machine) can decode. The Testing with Concept Activation Vectors framework quantifies how much a user-defined concept influences a prediction and can therefore be turned into an auxiliary loss that pushes a policy to maximise the \emph{explainability gradient} of its hidden units \citep{kim2018tcav}. Concept-Bottleneck Models extend this idea by forcing an intermediate layer to predict an explicit vector of human-labelled concepts; modifying that vector predictably steers the output \citep{koh2020cbm}. Sun \emph{et al.} have recently grafted the same bottleneck onto large language models (CB-LLM), enabling both controllable generation and faithful, inline explanations \citep{sun2024cbllm}. Dictionary-learning work at Anthropic decomposes LLM activations into thousands of nearly \emph{monosemantic} features, and shows that fine-tuning on a sparsity-plus-orthogonality objective can make those features more disentangled and hence more legible to outside probes—an intrinsic “make-me-understandable’’ pressure \citep{bricken2023monosemantic}. In multi-agent settings, Li \emph{et al.} ground emergent messages in a natural-language embedding space so that teammates (or humans) can decode them; agents are explicitly rewarded when a listener recovers the speaker’s latent goal from the transmitted symbol sequence \citep{li2024langground}. Bhardwaj pushes the idea further with Differentiable Inter-Agent Transformers: each agent’s self-attention heads are regularised so that the tokens they emit form a sparse, symbolic code that a partner model can reconstruct, effectively baking an \emph{interpretability-for-others} drive into the policy network itself \citep{bhardwaj2025diat}. Together, these techniques outline a concrete implementation path for the “wanting to be understood’’ motive: add an auxiliary reward (or architectural bottleneck) that maximises the mutual information between an agent’s private state and a public, human-decodable representation, while preserving task performance. \\

The recent advances in \emph{intrinsically interpretable} machine-learning architectures illustrate—in silico—the very causal chain we have posited for humans. When an auxiliary “be-understood’’ reward is added, a network is driven to package its opaque activations into a sparse, human-comprehensible code (Testing with Concept Activation Vectors, bottleneck vectors, monosemantic features, or emergent symbolic messages) \citep{kim2018tcav,koh2020cbm,bricken2023monosemantic,li2024langground,bhardwaj2025diat}. These externalised codes then become globally available to other agents and to the system’s own higher-level controllers, functioning exactly like the public representations—words, diagrams, gestures—that, in our story, bootstrapped human access consciousness from private phenomenology. At the same time, the codes remain a drastic compression of the underlying high-bandwidth state space (e.g.\ millions of parameters distilled into a few concept logits), mirroring our argument that no symbol can fully capture the sensorimotor manifold it stands for. Thus, the successes and limits of current LLM-interpretability work empirically vindicate the paper’s claims that a drive to be understood can be instantiated as a formal optimisation target.\\

\section{Conclusion}

To summarise: the very persistence of the hard problem is a clue to the distinctive, socially constructed form of consciousness that has emerged over hundreds of millennia of human cultural evolution. From infancy, humans work to make their ineffable experiences effable—first through proto-conversation and joint attention \citep{trevarthen1979communication,tomasello1995joint}, and later through increasingly sophisticated external representations: art \citep{dissanayake1995homo,noe2023entanglement}, myth and narrative \citep{boyd2005origin,harari2014sapiens}, science and philosophy \citep{donald1991origin}, and everyday folk-psychological storytelling \citep{hutto2009folk}. These practices expand the highest tiers of our sensorimotor-contingency hierarchy (Figure 1), allowing others to know, empathise, understand, and value our internal lives. In doing so, we construct a shared “common ground’’ of mutually accessible world-models—aligning beliefs, goals, and desires—and thereby create the mutual awareness that defines human social cognition \citep{malafouris2013things,clark2008supersizing}. \\

% The claim in this paper is that it is the intrinsic motivation for inventing new ways of achieving mutual awareness (top level SMCs) which makes us unique as a species, and which has allowed the bootstrapping of conscious (mutual) awareness. It has resulted in the invention of `consciousness technologies', namely external representations, that can be socially used for the creation of new kinds of mutual awareness \citep{fernando2024origin}. Our task then is to identify precisely how such a motivation for shared intensionality, primary and secondary intersubjectivity could be neurally and cognitively implemented \citep{tomasello2023differences} \citep{fernando2025wanting}, which if abstracted would motivate LLMs for example to produce emergent language \citep{lazaridou2020emergent} without already presupposing the existence of externally imposed language games \cite{wittgenstein1953philosophical}. \\

The claim in this paper is that a species-unique intrinsic motivation to invent ever-new means of achieving mutual awareness—the highest tier of sensorimotor contingencies—bootstrapped human conscious (mutual) awareness. This impulse has repeatedly yielded what Donald calls consciousness technologies—external representations such as gesture, writing, diagrams, and digital media that create new channels for shared experience \citep{donald1991origin,malafouris2013things,fernando2024origin}. Explaining that motivation requires specifying the neural and cognitive substrates of primary and secondary intersubjectivity: mirror-neuron and mentalising networks, second-person “interaction engines,’’ and predictive-coding circuits tuned for other-minds modelling \citep{frith2006social,schilbach2013toward,kilner2009predictive,tomasello2023differences,fernando2025wanting}. Abstracting such mechanisms could furnish large language models with intrinsic social drive, enabling truly emergent communication without pre-programmed language games \citep{lazaridou2020emergent,jaques2019social}, thereby moving AI a step closer to our biologically grounded quest for mutual understanding envisioned by \cite{wittgenstein1953}. \\

% We have focused on a socially driven source of the “explanatory gap.” Specifically, we argue that humans possess a unique, evolutionarily derived motivation to make our internal states mutually accessible, to “feel felt” and to minimize private discrepancies in how we experience the world. This leads to repeated attempts at externalizing or objectifying our subjective experience—attempts which inevitably fail to capture the full sensorimotor richness of what is being labeled. In illusionist terms \citep{frankish2016illusionism}, one might say that part of why our introspective limits feel like a special extra “ineffable” property is that we cannot exhaustively refer to that property, even though we are strongly motivated (socially and personally) to do so. From our perspective, therefore, the “why does it feel like anything at all?” question is inseparable from an inherently social puzzle: why do we care so urgently to pin down, share, and explain what we feel—and why is the inevitable failure of that project so persistently vexing?\\

We have focused on a socially driven source of the “explanatory gap.” Humans appear to possess an evolutionarily honed motive to make internal states mutually accessible—to have others “feel that we feel’’ \citep{siegel2012developing,fonagy2009mentalization}—and to minimise private discrepancies in how we experience the world \citep{frith2006social}. This compels us to externalise or objectify our subjectivity, even though every verbal or artistic label under-samples the sensorimotor richness it targets, as shown by overflow experiments in vision \citep{landman2003large,block2007consciousness}. In illusionist terms \citep{frankish2016illusionism}, our sense of ineffability arises because introspection delivers only coarse, theory-laden proxies for underlying experience, while social demands push us to treat those proxies as if they should be complete—a tension exposed by classic findings on the illusion of introspective insight \citep{nisbett1977telling}. \\

Our proposal does not resolve the ontological hard problem—why physical systems are accompanied by phenomenal experience at all. What we have offered is a meta-explanation of why creatures like us are perennially dissatisfied with every candidate physical story and thus keep re-raising the hard problem. In brief: a species-specific motivation to render private experience public collides with the severe compression imposed by language and other “consciousness technologies,” leaving an irreducible residue that feels ineffable.\\

If we could bring ourselves to really feel what we rationally can understand, that explanation is a norm-governed communicative act, the explanatory gap would collapse, but we cannot bring ourselves to see that in the case of explanations of our experience. Pragmatist and audience-centred accounts of science stress that an explanation succeeds when it positions hearers to ‘grasp’ a pattern, not when it reproduces the experience itself \cite{Potochnik2016}. Wittgenstein’s attack on private language shows that talk of sensations is already rule-bound and publicly checkable \cite{wittgenstein1953}, while Brandom’s inferentialism cashes this out in terms of socially policed reasons \cite{Brandom2000}. Speech-act theorists make this explicit: explaining is the illocution helping others to understand \cite{Gaszczyk2023}. Phenomenology reframes consciousness as embodied expression in a shared world \cite{MerleauPonty2012}, and enactivist work on participatory sense-making models subjectivity itself as an interactional achievement \cite{DeJaegher2007}. Against this backdrop, our demand that explanations also conjure the qualitative feel of experience looks misplaced: the explanation should have been complete when it earned its explanatory role in discourse. We propose it is our intrinsic motivation which seeks total and perfect communicative union with another which makes us behave so unreasonably in this regard. But that it is also this very motivation which has allowed us to invent our unique kind of external representation dependent conscious awareness. \\

\section{Acknowledgements}

Many thanks to David Amos for comments and help on the logic of the manuscript. Thanks to Emily Nicholls for discussions, advice and support during the writing of the manuscript. 

% Bibliography components
\bibliographystyle{abbrvnat}
\nobibliography*
\bibliography{template_refs}

\end{document}